\documentclass[letterpaper, 10 pt, conference]{ieeeconf}  

\IEEEoverridecommandlockouts                              

\usepackage[english]{babel}
\usepackage[utf8]{inputenc}

\usepackage{graphics} 
\usepackage{epsfig} 

\usepackage{amsmath,amssymb,amsfonts}
\usepackage{float}
\usepackage{graphicx}
\usepackage[colorlinks=true, allcolors=blue]{hyperref}
\usepackage{newtxtext, newtxmath}
\usepackage{cite}
\let\labelindent\relax
\usepackage[shortlabels]{enumitem}

\title{\LARGE \bf
Behavioral Inequalities
}

\author{Soutrik Bandyopadhyay$^{1}$, Debasattam Pal$^{2}$ and Shubhendu Bhasin$^{1}$
\thanks{$^{1}$Soutrik Bandyopadhyay and Shubhendu Bhasin are with the Department
  of Electrical Engineering, Indian Institute of Technology Delhi
        {\tt\small \{Soutrik.Bandyopadhyay,sbhasin\}@ee.iitd.ac.in}}%
\thanks{$^{2}$Debasattam Pal is with the Department of Electrical Engineering,
  Indian Institute of Technology Bombay
        {\tt\small debasattam@ee.iitb.ac.in}}%
}

\renewcommand{\Re}{\mathbb{R}}

\newcommand{\inprod}[2]{\left\langle #1, #2 \right\rangle}

\let\oldforall\forall
\renewcommand{\forall}{\; \oldforall \, }

\let\oldexist\exists
\renewcommand{\exists}{\; \oldexist \, }

\newtheorem{lem}{Lemma}
\newtheorem{thm}{Theorem}
\newtheorem{rem}{Remark}

\newtheorem{defn}{Definition}
\newtheorem{examp}{Example}

\DeclareMathOperator{\kernel}{ker}

\begin{document}

\maketitle
\thispagestyle{empty}
\pagestyle{empty}

\begin{abstract}
 We introduce behavioral inequalities as a way to model dynamical
 systems defined by inequalities among their variables of interest. We claim
 that such a formulation enables the representation of
 safety-aware dynamical systems, systems with bounds on disturbances, practical design limits and
 operational boundaries, etc. We develop a necessary and sufficient condition
 for the existence of solutions to such behavioral inequalities and provide a
 parametrization of solutions when they exist. Finally, we show the efficacy of the
 proposed method in two practical examples.
\end{abstract}


\section{Introduction}%

A fundamental way of analyzing dynamical systems is the behavioral approach
where a system is defined as a set of trajectories following some underlying
relation among variables. As opposed to the input-output structure of describing
systems, the behavioral approach relinquishes the idea of specifying pre-defined
models in favor of establishing relationships between observable trajectories.
Championed by Jan C. Willems
\cite{willems1986time1,willems1986time2,willems1987time3,willems1991paradigms,willems1997introduction} in 1980s,
the behavioral systems
theory has gained traction in recent years to solve problems in  data-driven
control, signal processing, robust control, and other areas (cf.
\cite{markovsky2021behavioral} and the references therein). Describing dynamical systems using the behavioral framework
affords us the generality to reason about systems from their observed
trajectories as opposed to explicit state-space formulations. Consider, for
example, the dynamics of the solar system
\cite{willems1991paradigms}, which may be complicated to define in terms of explicit state-space models; instead,
its definition is more intuitive as the set of all trajectories of the planets
and their moons that satisfy Kepler's laws of planetary motion. Such a
formulation yields a geometric view of the system and is often much simpler than
traditional frequency/state-space domain analysis \cite{markovsky2021behavioral}.

Behavioral approach has found reasonable success in the literature for
identifying system models from observed trajectory data \cite{roorda1995global,van2012subspace,markovsky2013software}.
As opposed to conventional system identification methods, these approaches are
non-parametric and can learn system behaviors directly from data. In addition to
that, direct data-driven control problem based on behavioral systems approach
was reported in \cite{markovsky2008data}. Subsequently, a class of algorithms were reported in the
literature which solve the predictive control problem using data. The so called
``Data-enabled predictive control (DeePC)''
\cite{coulson2019data,baros2022online,alpago2020extended} solves the optimal
control problem in a data-driven fashion.

In addition to methods cited above, behavioral approach has been adopted
to solve missing data estimation problems \cite{markovsky2016missing}, robust
control \cite{coulson2019regularized,coulson2021distributionally}, and
data-driven noisy estimation problems \cite{kailath2000linear} to name a few. We
refer the reader to \cite{markovsky2021behavioral} for a detailed literature review.

In the literature cited above, the behaviors are described using equations among
the variables of interest. However, a large number of systems of practical
importance are characterized by inequality constraints in the signals. For
example, in safety-critical systems \cite{ames2016TAC,ames2019ECC}, safety
constraints are typically expressed in terms of inequalities of
state/input/output. While the importance of modeling systems via inequalities
was briefly mentioned in \cite{willems1997introduction}, the literature lacks theoretical discussion and
mathematical development surrounding these behavioral inequalities.
The questions on the existence of solutions and parametrization of solutions
remain unanswered in the literature.




The contributions of this paper are four-fold. First, we introduce behavioral
inequalities, which model systems with inequality constraints on the
variables of interest. Subsequently, we discuss the existence of solutions for
such behavioral inequalities. Then, we develop the necessary and sufficient
conditions for the feasibility of behavioral inequalities in conjunction with
behavioral equations. When solutions to the behavioral inequalities exist,
we parameterize the solutions using generalized slack variables.
Finally, we provide practical examples to demonstrate the efficacy of the
proposed method.

The rest of the paper is organized as follows - Section \ref{sec:prelim}
provides a brief refresher on the behavioral approach to systems theory and
infinite dimensional vector spaces. Section \ref{sec:behav} introduces the core
idea of behavioral inequalities. We develop the necessary and sufficient
conditions for the existence of solutions for linear
behavioral inequalities in Section \ref{sec:existence}. Existence of solutions
for mixed equalities and inequalities is discussed in Section \ref{sec:mixed}.
In Section \ref{sec:param}, we discuss parametrization of solutions of
behavioral inequalities. Finally, Section \ref{sec:examples} provides practical
examples to demonstrate the efficacy of the proposed method.

\section{Preliminaries}%
\label{sec:prelim}
In this section we discuss mathematical tools pertinent to the
present work.
\subsection{Behavioral Systems Theory}%
In the context of behavioral systems theory, a dynamical system is defined as the tuple
\cite[Definition II.1]{willems1991paradigms}
\begin{equation}
 \Sigma \triangleq (\mathcal{T},\mathcal{W},\mathcal{B}),
\end{equation}
where $\mathcal{T}$ denotes the indexing set, $\mathcal{W}$ denotes the alphabet set, and the behavior
$\mathcal{B}$ denotes the set containing the trajectories of the system. In
other words,
$\mathcal{B} \subseteq \mathbb{U} \triangleq \{w: \mathcal{T} \to \mathcal{W}\}$.
In this notion of dynamical systems, the traditional distinction of inputs,
states and outputs are no longer present. Instead, a behavior is termed as a set
of trajectories satisfying a governing equation (or inequality in the context of
this paper).

For example, a discrete-time linear shift-invariant system can be represented
using the behavioral systems approach by defining
\begin{equation}
  \label{eq:linear_behavior}
  \mathcal{B} \triangleq \{w \in (\Re^{q})^{\mathbb{Z}} : R(\sigma,\sigma^{-1})w = 0\},
\end{equation}
where $R(\cdot) \in \mathbb{R}[\sigma,\sigma^{-1}]^{g \times q}$ is a polynomial matrix of size $g \times q$ in the indeterminates
$\sigma$ and $\sigma^{-1}$ with real coefficients. In this paper, we define the vector space
$(\Re^{q})^{\mathbb{Z}} \triangleq \{w: \mathbb{Z} \to \Re^{q}\}$. Here the indexing set
$\mathcal{T}= \mathbb{Z}$ and the alphabet set $\mathcal{W} = \Re^{q}$. For the
discrete-time case, the indeterminate $\sigma $ typically denotes the
shift operator, i.e.,
\begin{equation}
  (\sigma w)(k) \triangleq w(k+1) \forall k \in  \mathbb{Z},
\end{equation}
and the indeterminate $\sigma^{-1}$ is the inverse-shift operator
\begin{equation}
  (\sigma^{-1} w)(k) \triangleq w(k-1) \forall k \in  \mathbb{Z}.
\end{equation}
Without loss of generality, the matrix shift operator $R(\cdot)$ can be represented as
\begin{equation}
  R(\sigma,\sigma^{-1}) \triangleq \sum_{i=L_1}^{L_2} R_i \sigma^{i},
\end{equation}
where $R_i \in \Re^{g \times  q} \forall i$ and $L_1, L_2 \in \mathbb{Z}$ with
$L_1 \leq L_2$. It can be verified that the operator $R(\sigma,\sigma^{-1})$
is a bounded linear operator. The behavior $\mathcal{B}$ can be represented as
the kernel of the matrix shift operator $\mathcal{B} = \kernel R(\sigma,\sigma^{-1})$.

We like to highlight here that the kernel representation in
\eqref{eq:linear_behavior} is written in the language of equations. In this
work, we discuss systems expressed via inequalities. We argue that such an
extension can help solve numerous problems in control theory, economics, robotics, etc.
We will introduce the concept of behavioral inequalities in Section \ref{sec:behav}.

\subsection{Infinite Dimensional Vector Spaces}%
\label{subsec:label}

Let $\mathcal{X}$ be a real vector space, and
$\inprod{\cdot}{\cdot} : \mathcal{X} \times \mathcal{X} \to \Re$ be an inner product defined on $\mathcal{X} \times \mathcal{Y}$ (cf.
\cite{kreyszig1991introductory}). For any subspace $E \subseteq \mathcal{X}$, we use the notation
$E^{*}$ to denote the vector space of all linear functionals on $E$, termed as
the \textit{dual space}.
A subset $\mathcal{K} \subseteq \mathcal{X}$ is said to be convex iff
$\forall w_1, w_2 \in \mathcal{K}$, we have $\alpha w_1 + (1 - \alpha) w_2 \in \mathcal{K} \forall \alpha \in [0,1] $.
The convex set $\mathcal{P} \subseteq \mathcal{X}$ is said to be a convex cone iff
$w \in \mathcal{P} $ implies that $ \alpha w \in \mathcal{P} \forall \alpha \geq 0$.

Using the notion of convex cones, we can define the generalized inequalities for
infinite dimensional vector spaces as follows - consider the convex cone
$\mathcal{P}$, the ordering ``$\leq$'' (with respect to $\mathcal{P}$) is defined by
$x_1 \leq x_2$ iff $(x_2 - x_1) \in \mathcal{P}$ \cite{luenberger1997optimization}. The cone $\mathcal{P}$ is
subsequently called as the \textit{positive cone}.
For real valued vector spaces $\mathcal{X}$ and $\mathcal{Y}$ and a continuous linear operator $A: \mathcal{X} \to  \mathcal{Y}$, the adjoint
operator $A^{*}: \mathcal{Y} \to \mathcal{X}$ is a continuous linear operator satisfying
\begin{equation}
  \inprod{Ax}{y} = \inprod{x}{A^{*}y} \forall x \in \mathcal{X}, y \in \mathcal{Y}.
\end{equation}
In addition to the above, in this paper, we restrict our purview to cones
$\mathcal{P}$ which satisfy the following:
\begin{equation}
  \label{eq:assumption}
  \inprod{x}{y} \geq 0 \forall x,y \in \mathcal{P}.
\end{equation}
This restriction will be used in the proof of Lemma \ref{lem:farkas}. One
example of such a cone for the vector space $(\Re^{q})^{\mathbb{Z}}$ is the
positive orthant
$\mathcal{E} \triangleq \{w \in (\Re^{q})^{\mathbb{Z}} | w_{i}(k) \geq 0 \forall k \in \mathbb{Z}, i \in \{1,2, \dots, q\}\}$.
For the vector space $(\Re^{q})^{\mathbb{Z}}$, we define the
inner product
\begin{equation}
  \inprod{x}{y} \triangleq \sum_{k \in \mathbb{Z}} x(k)^{\mathrm{T}}y(k) \forall x,y \in (\Re^{q})^{\mathbb{Z}},
\end{equation}
where the set $\{k \in \mathbb{Z} \; | \; x(k)^{\mathrm{T}}y(k) \neq 0\}$ is finite.

\section{Behavioral Inequalities}
\label{sec:behav}
We now introduce a class of behavioral models
defined by inequalities.

\begin{defn}[Behavioral Inequalities]
  \label{defn:behavioral_ineq}
Consider the universum $\mathbb{U}$, a vector space $\mathbb{E}$ and
$f: \mathbb{U}\to \mathbb{E}$. The behavioral model $(\mathbb{U},\mathcal{B})$
with $\mathcal{B} \triangleq \{w \in \mathbb{U} | f(w) \leq 0 \}$ is called a
behavioral inequality, where the ordering ``$\leq$'' is with respect to the
positive cone $\mathcal{P} \subseteq \mathbb{E}$.
\end{defn}

Definition \ref{defn:behavioral_ineq} admits a wide variety of mathematical
models of practical significance. Let us now look at
temporal inequalities which are affine in the variable of interest as follows:
\begin{equation}
  a_{L}w(k+L) + a_{L-1}w(k+L-1) + \dots + a_{0}w(k) \leq b \forall k \in \mathbb{Z},
\end{equation}
where $w \in (\mathbb{R})^{\mathbb{Z}}$,
$a_{i},b \in \mathbb{R} \forall i \in \{0,1, \dots, L\}$ and $L \in \mathbb{N}$.
We can write this inequality utilizing the shift operator as
\begin{equation}
  \mathbf{a}(\sigma,\sigma^{-1}) w \leq \mathbf{b},
\end{equation}
where $\mathbf{a} \in \mathbb{R}[\sigma,\sigma^{-1}]$ is a polynomial in the indeterminates
$\sigma $ and $\sigma^{-1}$ with real coefficients,
$\mathbf{b} \triangleq \{ w \in (\mathbb{R})^{\mathbb{Z}} | w(k) = b \forall k \in \mathbb{Z} \}$
is the constant trajectory with the value of $b$ at each time instant
and ``$\leq$'' operator is defined w.r.t. the positive cone $\mathcal{P} \subset (\mathbb{R})^{\mathbb{Z}}$.
These inequality constraints can encode safety-critical conditions, bounds on
disturbances, actuation limits, stability conditions in control applications to
name a few.
\begin{examp}
  Consider the behavioral inequality
  \begin{equation}
    \label{eq:baccha_example}
    (\sigma^{2} - \sigma + 1) w \leq 2,
  \end{equation}
where $w \in (\mathbb{R})^{\mathbb{N}}$. Let the initial conditions of $w$ be
$w(1) = 1$, $w(2) = 1$. We can verify that $w = \{1,1,1,1, \dots \}$,
$w = \{1,1,1.5,2, 2.5, \dots \}$ and $w = \{1,1,0.5,0,0,0, \dots \}$ are solutions to
\eqref{eq:baccha_example}. As opposed to behavioral equations, behavioral
inequalities may have infinite solutions for a particular initial condition.
\end{examp}

In general, we may write a system of temporal inequalities with constant
coefficients as
\begin{equation}
  \label{eq:general_ineq}
 H(\sigma,\sigma^{-1})w \leq g,
\end{equation}
where $w \in (\Re^{q})^{\mathbb{Z}}$, $H\in \mathbb{R}[\sigma,\sigma^{-1}]^{l \times q}$ and
$g \in (\Re^{l})^{\mathbb{Z}}$. The operator ``$\leq $'' is defined with respect to the
positive cone $\mathcal{P} \subset (\Re^{l})^{\mathbb{Z}}$. We now define the
behavioral inequality $\mathcal{B}_{in}$ as
\begin{equation}
  \label{eq:behav_defn}
 \mathcal{B}_{in} \triangleq \{w \in (\Re^{q})^{\mathbb{Z}} | H(\sigma,\sigma^{-1})w \leq g \}.
\end{equation}
In other words, a trajectory $w$ is said to be a solution (if any) to the behavioral inequality
\eqref{eq:general_ineq} if the transformation $k = H(\sigma,\sigma^{-1})w$ satisfies $k \leq g$
(i.e., $g - k \in \mathcal{P}$).

\section{Feasibility of Behavioral Inequalities (Existence of Solutions)}%
\label{sec:existence}

We now study the existence of solutions for the behavioral inequality in
\eqref{eq:behav_defn}. In order to derive the necessary and sufficient
conditions for feasibility of behavioral inequalities, we now state the
generalization of the theorem of alternatives \cite{boyd2004convex}, i.e., the
Generalized Farkas' Lemma \cite{clark2006necessary}.

\begin{lem}[Generalized Farkas' Lemma \cite{clark2006necessary}]
  \label{lem:farkas}
  Consider two real valued vector spaces $\mathcal{X}$, $\mathcal{Y}$ with an inner product
  $\inprod{\cdot}{\cdot}$ defined on $\mathcal{Y}$. Let $A: \mathcal{X} \to \mathcal{Y}$ be a linear operator and
  $b \in  \mathcal{Y}$ be a constant vector. Let the vector space $\mathcal{Y}$
  be ordered by the operator ``$\leq $'' w.r.t. the positive cone
  $\mathcal{P} \subset \mathcal{Y}$ satisfying \eqref{eq:assumption}.
  Then, exactly one of the following statements is true
  \begin{enumerate}[label={(\textbf{\Alph*})}]
    \item $\exists x \in \mathcal{X}$ such that $Ax \leq b$
    \item $\exists y \geq 0$ such that
            $y \in \kernel{A^{*}}$ and $\inprod{y}{b} < 0$,
          where $A^{*}$ is the adjoint of $A$.
  \end{enumerate}
 \begin{proof}
   The proof consists of 3 parts
   \begin{enumerate}
     \item Let (\textbf{A}) be true. Thus, $\exists x \in \mathcal{X}$ such that
           $Ax \leq b$. Then for any $y \in \mathcal{Y}$ with $y\geq 0$, we can compute the inner product
           \begin{equation}
             \inprod{y}{Ax} \leq \inprod{y}{b}.
           \end{equation}
           Assume that $y \in \kernel{A^{*}}$. However,
           using the definition of the adjoint operator, we obtain
           \begin{equation}
             \inprod{y}{Ax} = \inprod{A^{*}y}{x} = 0 \leq \inprod{y}{b},
           \end{equation}
           which contradicts $\inprod{y}{b} < 0$, which is the condition of (\textbf{B}). Thus feasibility of (\textbf{A})
           implies (\textbf{B}) is false.
     \item Now, let (\textbf{B}) be true. Thus $\exists y \in \mathcal{Y}$ with $y\geq 0$
        such that $y \in \kernel{A^{*}}$ and $\inprod{y}{b} < 0$.
           Assume, for contradiction, $\exists x$ such that $Ax \leq b$. Again we write
           using $y \geq 0$
           \begin{equation}
             \inprod{y}{Ax} \leq \inprod{y}{b}.
           \end{equation}
           Using properties of adjoint operator, we write
           \begin{equation}
             \inprod{y}{Ax} = \inprod{A^{*}y}{x} = 0 \leq \inprod{y}{b},
           \end{equation}
           which again contradicts $\inprod{y}{b} < 0$. Thus if (\textbf{B}) is true, (\textbf{A})
           is false.
     \item We now show that both (\textbf{A}) and (\textbf{B}) can't simultaneously be false.
           Let (\textbf{A}) be false. Then it can be shown that the sets
           $\{Ax | x \in \mathcal{X}\}$ and
           $\{b - v | \forall v \in \mathcal{X}, v \geq  0  \}$ are disjoint (do
           not intersect). By using the Hahn-Banach Separation theorem \cite{luenberger1997optimization}, there exists a
           separating hyperplane that partitions the vector space $\mathcal{Y}$
           such that the following strict inequality holds
           \begin{equation}
             \inprod{y}{Ax} > \inprod{y}{b-v} \forall x \in \mathcal{X}.
           \end{equation}
           For an appropriate choice of $v$, we can obtain a $y \geq 0$ such that
           $y \in \kernel{A^{*}}$. Thus
           \begin{equation}
             \inprod{y}{Ax} = \inprod{A^{*}y}{x} = 0 > \inprod{y}{b} \forall  x \in \mathcal{X},
           \end{equation}
           which is exactly the condition for statement (\textbf{B}). Thus if (\textbf{A}) is
           false, then (\textbf{B}) must be true and vice versa.
   \end{enumerate}
   Thus, exactly one of the statements (\textbf{A}) and (\textbf{B}) must be true.
 \end{proof}
\end{lem}

Lemma \ref{lem:farkas} can be used to derive a necessary and sufficient condition for the
feasibility of behavioral inequality in \eqref{eq:behav_defn}. In order to do that, we first derive
the adjoint of the polynomial shift operator in the following Lemma.

\begin{lem}[Adjoint of polynomial shift operator]
  \label{lem:adjoint}
For a polynomial shift operator $R(\sigma,\sigma^{-1}) \in \Re[\sigma,\sigma^{-1}]^{l \times q}$, the corresponding adjoint
operator is $R^{\mathrm{T}}(\sigma^{-1},\sigma)$, where $\sigma ^{-1}$ denotes the inverse shift
operator defined as $(\sigma^{-1}w)(k) \triangleq w(k-1) \forall k \in \mathbb{Z}$.
\begin{proof}
  Consider the shift operator $\sigma $, the adjoint operator is defined as

  \begin{equation}
    \inprod{\sigma w}{y^{*}} =
    \inprod{w}{\sigma^{*}y^{*}} \forall w,
  \end{equation}
    where $\sigma^{*}$ is the adjoint operator for the shift operator.
  Using the definition of inner product, we can write the left-hand side of the
  equation as
 \begin{equation}
  \sum_{k\in \mathbb{Z}} w(k+1)^{\mathrm{T}} y^{*}(k).
 \end{equation}
 Shifting the summation one step, we have
 \begin{equation}
   \sum_{l\in \mathbb{Z}} w(l)^{\mathrm{T}} y^{*}(l - 1)
   = \inprod{w}{\sigma^{-1}y^{*}}.
 \end{equation}
 Thus the adjoint of the shift operator ($\sigma$) is the inverse-shift
 operator ($\sigma^{-1}$).

 Now, without loss of generality, we can write
 \begin{equation}
   R(\sigma,\sigma^{-1}) = \sum_{k=N_1}^{N_2} R_k \sigma^{k},
 \end{equation}
where $N_1,N_2 \in \mathbb{N}$ with $N_1 \leq  N_2$. Using the properties of adjoint, we write
\begin{equation}
\begin{aligned}
  R(\sigma,\sigma^{-1})^{*} &= \sum_{k=N_1}^{N_2} (R_k \sigma^{k})^{*}
  = \sum_{k=N_1}^{N_2} R_k^{\mathrm{T}} \sigma^{-k} = R^{\mathrm{T}}(\sigma^{-1},\sigma),
\end{aligned}
\end{equation}
which completes the proof.
\end{proof}
\end{lem}

We now write the necessary and sufficient condition for feasibility of the
behavioral inequality in \eqref{eq:behav_defn}.

\begin{thm}[Feasibility condition for behavioral inequality]
  \label{thm:main}
 Given the behavioral inequality defined by
 \begin{equation}
  \mathcal{B}_{in} \triangleq \{w \in (\Re^{q})^{\mathbb{Z}} | H(\sigma,\sigma^{-1})w \leq  g\},
 \end{equation}
 the behavior set is non-empty iff there does not exist a
 $y \in \kernel{H^{\mathrm{T}}(\sigma^{-1},\sigma)}$ with $y\geq 0$ and $\inprod{y}{g} < 0$.

\begin{proof}
The proof follows from the trivial application of Lemma \ref{lem:adjoint} in
Lemma \ref{lem:farkas}.
\end{proof}

\end{thm}

Using Theorem \ref{thm:main}, we can check for the existence of solutions for a
given
behavioral inequality.

\begin{examp}
  \label{examp:feasible_ineq}
Consider the behavioral inequality given by
\begin{equation}
  \label{eq:simple_feasible_examp}
  \begin{bmatrix}
    \sigma +1 & 1 \\
    1 & \sigma
  \end{bmatrix}
  w
  \leq
  \begin{bmatrix}
    15 \\
    10
  \end{bmatrix},
\end{equation}
where $w \in (\Re^{2})^{\mathbb{Z}}$. Now, to check for the existence of solutions for
the inequality mentioned above, we consider the kernel of the adjoint of the
matrix shift operator as
\begin{equation}
  \begin{bmatrix}
    \sigma^{-1} +1 & 1 \\
    1 & \sigma^{-1}
  \end{bmatrix} y = 0,
\end{equation}
where $y \in (\Re^{2})^{\mathbb{Z}}$. Now, performing unimodular row operations
(polynomial matrix operations whose determinant is a unit in the operator ring
\cite{willems1997introduction}) we obtain

\begin{equation}
  \begin{bmatrix}
    1 & \sigma^{-1} \\
    0 & 1 - \sigma^{-1} - \sigma^{-2}
  \end{bmatrix} y = 0.
\end{equation}
Subsequently, we write
\begin{equation}
\begin{aligned}
  y_1(k) + y_2(k-1) &= 0, \\
  y_2(k) - y_2(k-1) - y_2(k-2) &= 0 \forall k \in \mathbb{Z}.
\end{aligned}
\end{equation}
It is easy to note that since $y_1(k) \geq 0$ and
$y_2(k) \geq 0 \forall k \in  \mathbb{Z}$, the above condition implies that
$y_1(k) = y_{2}(k) = 0 \forall k \in \mathbb{Z}$. Subsequently,
$\inprod{y}{g} = 0$ which implies that \eqref{eq:simple_feasible_examp} is
feasible using Theorem \ref{thm:main}. One can easily verify this fact by
observing that the trajectory $w_1(k) = w_2(k) = 0 \forall k \in \mathbb{Z}$ satisfies
\eqref{eq:simple_feasible_examp}.

\end{examp}


\section{Mixed Behavioral Inequalities and Equations}%
\label{sec:mixed}
In practical systems, trajectories may involve both equality and inequality
constraints on the variables. Until now, we have discussed behavioral
inequalities. We now delve into systems where behavioral inequalities appear in
conjunction with behavioral equalities. Consider the behavioral system as
follows
\begin{equation}
  \label{eq:mixed_behav}
  \mathcal{B} \triangleq \{w \in (\Re^{q})^{\mathbb{Z}} |
  R(\sigma,\sigma^{-1})w = d,
  H(\sigma,\sigma^{-1})w \leq g
  \},
\end{equation}
where $R(\sigma,\sigma^{-1}) \in \mathbb{R}[\sigma,\sigma^{-1}]^{l_{e}\times q}$,
$H(\sigma,\sigma^{-1}) \in \mathbb{R}[\sigma,\sigma^{-1}]^{l_{i}\times q}$, $d \in (\Re^{l_{e}})^{\mathbb{Z}}$
and
$g \in (\Re^{l_{i}})^{\mathbb{Z}}$. We now state the Theorem for the feasibility
of \eqref{eq:mixed_behav}.

\begin{thm}
\label{thm:mixed_feasibility}
  Given the behavioral system in \eqref{eq:mixed_behav}, $\mathcal{B}$ is
  non-empty if and only if there does not exist
  $y \in (\Re^{l_{e}})^{\mathbb{Z}}$, $z \in (\Re^{l_{i}})^{\mathbb{Z}}$
  such that $z\geq 0$,
  \begin{equation}
    \begin{bmatrix}
      y \\
      z
    \end{bmatrix} \in
    \kernel{
      \begin{bmatrix}
        R^{\mathrm{T}}(\sigma^{-1},\sigma) &
        H^{\mathrm{T}}(\sigma^{-1},\sigma)
      \end{bmatrix}
    },
  \end{equation}
  and
  \begin{equation}
    \inprod{y}{d} + \inprod{z}{g} < 0.
  \end{equation}

\end{thm}
\begin{proof}
  We rewrite the mixed behavioral system in \eqref{eq:mixed_behav} as
  \begin{equation}
    \label{eq:feasiblity_augment}
    \underbrace{
    \begin{bmatrix}
      R(\sigma,\sigma^{-1})\\
      -R(\sigma,\sigma^{-1})\\
      H(\sigma,\sigma^{-1})
    \end{bmatrix}
    }_{\triangleq H'(\sigma,\sigma^{-1})}
    w
    \leq
    \underbrace{
    \begin{bmatrix}
      d \\
      -d \\
      g
    \end{bmatrix}}_{\triangleq g'}
  ,
  \end{equation}
  where the augmented shift operator
  $H'(\sigma,\sigma^{-1}) \in \mathbb{R}[\sigma,\sigma^{-1}]^{(2l_{e} + l_{i}) \times q}$ and
  $g' \in (\Re^{(2l_{e} + l_{i})})^{\mathbb{Z}}$.
  Now we can apply Theorem \ref{thm:main} to this augmented system to obtain the
  necessary and sufficient condition for the feasibility of
  \eqref{eq:feasiblity_augment} and consequently of \eqref{eq:mixed_behav}.
  Consider $\lambda \in  (\Re^{(2l_{e} + l_{i})})^{\mathbb{Z}}$ such that
  $\lambda \geq 0$ and $\lambda \in \kernel{H'(\sigma,\sigma^{-1})^{*}}$. Then the negative condition for Theorem
  \ref{thm:main} becomes $\inprod{\lambda }{g'} < 0$. We can partition $\lambda $ as $[y_1^{\mathrm{T}}, y_2^{\mathrm{T}}, z^{\mathrm{T}}]^{\mathrm{T}}$ satisfying
  \begin{equation}
    \begin{bmatrix}
      R^{\mathrm{T}}(\sigma^{-1},\sigma) &
      -R^{\mathrm{T}}(\sigma^{-1},\sigma) &
      H^{\mathrm{T}}(\sigma^{-1},\sigma)
    \end{bmatrix}
    \begin{bmatrix}
      y_1 \\
      y_2 \\
      z
    \end{bmatrix} = 0,
  \end{equation}
where $y_1, y_2 \in (\Re^{l_{e}})^{\mathbb{Z}}$ and $z \in (\Re^{l_{i}})^{\mathbb{Z}} $.
Similarly the inner product
$\inprod{\lambda }{g'} = \inprod{y_1}{d} + \inprod{y_2}{-d} + \inprod{z}{g}$. Defining
$y \triangleq y_1 - y_2$, we have the condition
  \begin{equation}
    \begin{bmatrix}
      R^{\mathrm{T}}(\sigma^{-1},\sigma) &
      H^{\mathrm{T}}(\sigma^{-1},\sigma)
    \end{bmatrix}
    \begin{bmatrix}
      y \\
      z
    \end{bmatrix} = 0,
  \end{equation}
and by linearity of the inner product $\inprod{y}{d} + \inprod{z}{g} < 0$.
Invoking Theorem \ref{thm:main}
completes the proof.
\end{proof}

\begin{rem}
Notice that for the case of mixed equalities and inequalities, the dual variable
corresponding to the behavioral equality need not be positive.
\end{rem}

We discuss an example of mixed behavioral inequalities and equalities in
Section \ref{sec:examples}.

\section{Characterization of Solutions of the Behavioral Inequalities}
\label{sec:param}

Provided a given set of behavioral inequalities is feasible, we now characterize
the set of solutions to the behavioral inequality in \eqref{eq:behav_defn}. In
order for this, we reformulate \eqref{eq:behav_defn} as
\begin{equation}
  \label{eq:slack_defn}
  \mathcal{B}_{in}
  =
  \{
  w \in (\Re^{q})^{\mathbb{Z}} |
  \exists s \in (\Re^{l})^{\mathbb{Z}}, s \geq 0,
  H(\sigma,\sigma^{-1})w + s = g
  \},
\end{equation}
where the auxiliary variable $s \in (\Re^{l})^{\mathbb{Z}}$ is called the slack variable
in the context of constrained optimization
\cite{luenberger1997optimization,boyd2004convex}.
We can write the condition in \eqref{eq:slack_defn} succinctly as
\begin{equation}
  \label{eq:augmented_slack}
  \underbrace{
  \begin{bmatrix}
    H(\sigma,\sigma^{-1}) & \mathbb{I}_{l}
  \end{bmatrix}
  }_{\triangleq H_{s}(\sigma,\sigma^{-1})}
  \begin{bmatrix}
    w \\
    s
  \end{bmatrix}
  = g
  ,
  \; \; s \geq 0,
\end{equation}
where the augmented shift operator $H_{s}(\sigma,\sigma^{-1}) \in \mathbb{R}[\sigma,\sigma^{-1}]^{l \times (q + l)}$
dictates the behavior of the augmented system $(w, s)$ and $\mathbb{I}_l$ denotes the $l \times l$ identity matrix.

Now, we can reduce the order of the augmented system in
\eqref{eq:augmented_slack} by unimodular row operations
\cite{willems1991paradigms}. Specifically, there exists a unimodular matrix
$U(\sigma,\sigma^{-1}) \in \mathbb{R}[\sigma,\sigma^{-1}]^{l \times l}$ such that
\begin{equation}
  H_{s}'(\sigma,\sigma^{-1}) \triangleq U(\sigma,\sigma^{-1}) H_{s}(\sigma,\sigma^{-1}),
\end{equation}
is in the upper-triangular form
\cite{willems1997introduction}. Thus we can rewrite \eqref{eq:augmented_slack}
in the equivalent form as
\begin{equation}
  H_{s}'(\sigma,\sigma^{-1}) \begin{bmatrix}
    w \\ s
  \end{bmatrix} = U(1) g, \;\; s\geq  0.
\end{equation}
Given an initial condition $w_0$, one can obtain a series of solutions for $w$
with respect to the trajectory of $s$. We can thus parametrize the solution $w$
of the behavioral inequality by $w(s) : \mathcal{S} \to (\Re^{q})^{\mathbb{Z}}$,
where
$\mathcal{S} \triangleq \{s \in (\Re^{l})^{\mathbb{Z}}, s\geq 0 | \exists w \in  (\Re^{q})^{\mathbb{Z}} , H_{s}'(\sigma,\sigma^{-1})[w^{\mathrm{T}} s^{\mathrm{T}}]^{\mathrm{T}} = g\}$
denotes the set of all possible slack trajectories.

\begin{examp}
  Consider the feasible behavioral inequality in Example
  \ref{examp:feasible_ineq}. We can rewrite the same as
  \begin{equation}
  \begin{bmatrix}
    \sigma +1 & 1 & 1 & 0\\
    1 & \sigma & 0 & 1
  \end{bmatrix}
  \begin{bmatrix}
    w_1 \\ w_2 \\ s_1 \\ s_2
  \end{bmatrix}
  =
  \begin{bmatrix}
    15 \\
    10
  \end{bmatrix}, \; s\geq 0.
  \end{equation}
  We can write the equivalent behavior by performing unimodular row operations
  \begin{equation}
    \label{eq:temporary}
  \begin{bmatrix}
    1 & \sigma  & 0 & 1\\
    0 & 1-\sigma -\sigma^2 & 1 & -1 -\sigma
  \end{bmatrix}
  \begin{bmatrix}
    w_1 \\ w_2 \\ s_1 \\ s_2
  \end{bmatrix}
  =
  \begin{bmatrix}
    10 \\
    -5
  \end{bmatrix}, \; s\geq 0.
  \end{equation}
  Observe that there are two pivots for the above polynomial matrix. Thus the
  rank of the augmented matrix is two, and consequently, we can choose the
  trajectories of $s_1$ and $s_2$ independently. We can expand
  \eqref{eq:temporary} to obtain the recursive relation
  \begin{equation}
    \label{eq:recursive}
\begin{aligned}
    w_1(k) + w_2(k+1) + s_2(k) &= 10, \\
  w_2(k) - w_2(k+1) -w_2(k+2)
  + s_1(k) &
             - s_2(k) \\
  - s_2(k+1) = -5 \forall k \in \mathbb{Z}.
\end{aligned}
  \end{equation}
  Given a set of initial conditions, one can choose a trajectory $s\geq 0$ to
  obtain the trajectory for $w$ by solving the recursive relation \eqref{eq:recursive}.
\end{examp}








\section{Practical Examples}%
\label{sec:examples}

\subsection{Safety-aware Dynamical Systems}%
Consider the discrete time linear time-invariant (LTI) system
\begin{equation}
\begin{aligned}
  x(k+1) &= Ax(k) + Bu(k), \\
  y(k) &= C x(k) + D u(k) \forall k \in \mathbb{Z},
\end{aligned}
\end{equation}
where
$x \in (\Re^{n})^{\mathbb{Z}}, u \in (\Re^{m})^{\mathbb{Z}}, y \in (\Re^{p})^{\mathbb{Z}}$
denote the state, action, and output variables, respectively and the matrices
$A, B, C,$ and $D$ have the appropriate dimensions. The system can be written in the
behavioral approach as

\begin{equation}
  \underbrace{
  \begin{bmatrix}
    \sigma \mathbb{I}_n - A  & -B & 0 \\
    -C  & -D & \mathbb{I}_{p} \\
  \end{bmatrix}}_{\triangleq R(\sigma,\sigma^{-1})}
  w = 0,
\end{equation}
where
$w(k) \triangleq [x(k)^{\mathrm{T}}, u(k)^{\mathrm{T}}, y(k)^{\mathrm{T}}]^{\mathrm{T}} \forall k \in \mathbb{Z}$
and $R(\sigma,\sigma^{-1}) \in \mathbb{R}[\sigma,\sigma^{-1}]^{(n+p) \times (n+m+p)}$. Here $\mathbb{I}_n$ and $\mathbb{I}_p$ denote the $n \times n$ and $p \times p$ identity matrices respectively.
In safety-aware settings in the real world, the dynamical systems must adhere to
user-defined constraints, which may include constraints on states, inputs,
outputs and their rates of change. In the framework proposed in this work, we
can easily write polytopic constraints on states, input, output, and rate of
change of input as
\begin{equation}
 \begin{bmatrix}
   H_{x} & 0                  & 0 \\
   0     & H_{u}              & 0 \\
   0     & 0                  & H_{y} \\
   0     & (\sigma -1) H_{du} & 0
 \end{bmatrix}
 \begin{bmatrix}
   g_{x}\\
   g_{u}\\
   g_{y}\\
   g_{du}\\
 \end{bmatrix}.
\end{equation}
 Notice that the safety-aware dynamical
system is of the form
\begin{equation}
  \mathcal{B}_{safe} \triangleq \{w\in (\mathbb{R}^{(n+m+p)})^{\mathbb{Z}} | R(\cdot)w = 0, H(\cdot)w \leq g \},
\end{equation}
which is a behavioral system with both equality and inequality constraints. We
can check the existence of solutions of the behavioral system as discussed in
sections above.

\begin{examp}
  \label{examp:infeasible_lti}
  Given the LTI system
  \begin{equation}
    \label{eq:safety_state}
    x(k+1) =
    \begin{bmatrix}
      2 & 0\\
      1 & -1
    \end{bmatrix}
    x(k)
    +
    \begin{bmatrix}
      0 \\ 1
    \end{bmatrix}
    u(k) \forall k \in \mathbb{Z},
  \end{equation}
  and let the state and actuation constraints be
  \begin{equation}
    \label{eq:safety_constraint}
   \begin{aligned}
     1 &\leq x_1(k) \leq 5, \\
     -5 &\leq x_2(k) \leq 5, \\
     -1 &\leq u(k) \leq 1 \forall k \in \mathbb{Z}.
    \end{aligned}
  \end{equation}
  Following the discussion above, we can write the constrained system by
  \begin{align}
    R(\sigma,\sigma^{-1}) &\triangleq
    \begin{bmatrix}
      \sigma - 2 & 0 & 0 \\
      - 1 & \sigma +1 & -1 \\
    \end{bmatrix}, \\
    H(\sigma,\sigma^{-1}) &\triangleq
    \begin{bmatrix}
      1  & 0  & 0 \\
      -1 & 0  & 0 \\
      0  & 1  & 0 \\
      0  & -1 & 0 \\
      0  & 0  & 1 \\
      0  & 0  & -1 \\
    \end{bmatrix},\\
    g &\triangleq
    \begin{bmatrix}
      5 &
      -1 &
      5 &
      5 &
      1 &
      1
    \end{bmatrix}^{\mathrm{T}}.
  \end{align}

  Using Theorem \ref{thm:mixed_feasibility}, we can test for the feasibility of
  the mixed behavioral system. Consider the kernel of the adjoint of the
  augmented matrix
  \begin{equation}
    \begin{bmatrix}
      R^{\mathrm{T}}(\sigma^{-1},\sigma) &
      H^{\mathrm{T}}(\sigma^{-1},\sigma)
    \end{bmatrix}
    \begin{bmatrix}
      y \\ z
    \end{bmatrix} = 0,
  \end{equation}
  where $y \in (\Re^{2})^{\mathbb{Z}}$ and $z \in (\Re^{6})^{\mathbb{Z}}$. Now we can
  perform elementary unimodular row operations to yield the upper triangular form
  \begin{equation*}
    \begin{bmatrix}
      \sigma^{-1} - 2 & -1 & 1 & -1 & 0 & 0 & 0  & 0 \\
      0                & 1  & 0 & 0  & 0 & 0 & -1 & 1 \\
      0                & 0  & 0 & 0  & 1 & -1 & \sigma^{-1} + 1 & -\sigma^{-1} - 1 \\
    \end{bmatrix}.
  \end{equation*}

  Notice that the rank of this matrix is 3, and thus by the rank-nullity theorem, we
  can freely choose the trajectories of 5 variables-
  $z_1$, $z_2$, $z_4$, $z_5$ and $z_6$.
  We can write the recursive relation
  \begin{equation}
   \begin{aligned}
     &y_1(k-1) - 2y_1(k) - y_2(k) + z_1(k) - z_2(k) = 0, \\
     &y_2(k) - z_5(k) + z_6(k) = 0, \\
     &z_3(k) - z_4(k) + z_5(k-1) + z_5(k)\\
     &- z_6(k-1) - z_6(k) = 0 \forall k \in  \mathbb{Z}.
    \end{aligned}
  \end{equation}

  We observe that by choosing $z \geq 0$, such that $z_1(k) = 1 \forall k \in \mathbb{Z}$,
  $z_2(k) = 10 \forall k \in \mathbb{Z}$ and setting
  $z_3(k) = z_4(k) = z_5(k) = z_6(k) = 0$, we obtain
  $\inprod{y}{0} + \inprod{z}{g} < 0$. Thus by using Theorem
  \ref{thm:mixed_feasibility}, the system \eqref{eq:safety_state} with the
  constraint \eqref{eq:safety_constraint} is infeasible. This can be further
  verified by observing that the control action has no influence on the state
  $x_1$ and the eigenvalue of the $x_1$ dynamics is $2$. Thus the constraints
  are infeasible. We demonstrate the same in Figure \ref{fig}.

 \begin{figure}[htpb]
   \centering
 \includegraphics[width=\linewidth]{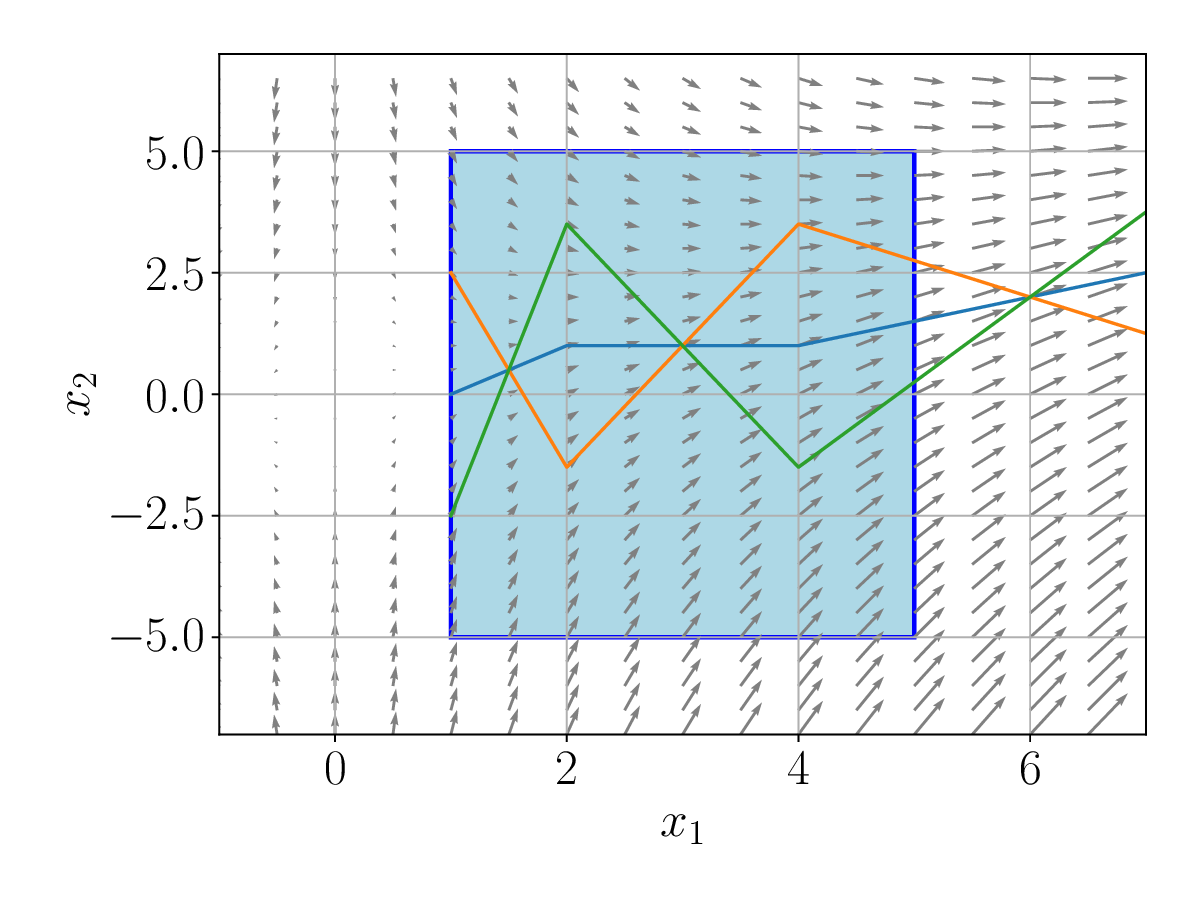}
 \caption{\label{fig} Infeasibility of behavioral inequality constraints
   discussed in Example \ref{examp:infeasible_lti}. The blue region shows the
   state constraints. The quiver plot for the dynamics of the system is shown in
 gray. We observe that trajectories originating in the constraint region
 eventually violate the safety constraint.}
 \end{figure}

\end{examp}

\subsection{Dynamic Inventory Management Problem}%
Consider a warehouse that needs to minimize long-term costs  while maintaining a
stock of goods to satisfy consumer demand. At each instant $k$, let $u(k)$, $x(k)$ and
$d(k)$ denote the goods ordered, goods in stock and the demand respectively. The
objective is to minimize
\begin{equation}
  \label{eq:cost}
  J \triangleq \sum_{k \in \mathbb{Z}} c_k u(k),
\end{equation}
where $c_{k}$ is the per-unit cost of procuring the goods at $k$-th instant. The
warehouse system is governed by the following constraints
\begin{align}
  x(k+1) &= x(k) + u(k) - d(k), \label{subeq:dyn}\\
  x(k) &+ u(k) \geq d(k), \label{subeq:dem}\\
  x(k),&\; u(k),\; d(k) \geq 0, \forall k \in \mathbb{Z}, \label{subeq:posi}
\end{align}
where \eqref{subeq:dyn} model the warehouse dynamics, \eqref{subeq:dem} denotes
the demand constraint and \eqref{subeq:posi} are the non-negativity constraints.

We can model the warehouse system using the proposed framework by defining
$w \triangleq [x^{\mathrm{T}}, u^{\mathrm{T}}, d^{\mathrm{T}}]^{\mathrm{T}} $ and writing
\begin{align}
\begin{bmatrix}
    (\sigma - 1) & -1 & -1
\end{bmatrix} w &= 0, \\
  \begin{bmatrix}
    -1 & -1 & 1 \\
    -1 & 0 & 0 \\
    0 & -1 & 0 \\
    0 & 0 & -1 \\
  \end{bmatrix} w &\leq 0.
\end{align}
Notice that the system above is a combination of behavioral equations and
inequalities. Using Theorem \ref{thm:mixed_feasibility}, it can be shown that
the above system is feasible. Subsequently, we can characterize the solutions
for the behavioral system above by utilizing generalized slack variables as
discussed in Section \ref{sec:param}. Depending upon the initial conditions, the
behavioral system may have a continuum of solutions. The optimal trajectory is
the one that minimizes \eqref{eq:cost}.



\section{Concluding Remarks}
In this paper, we proposed ``behavioral inequalities'' as a way to model
dynamical systems which are defined using inequalities among variables of
interest. We claim that such a formulation affords us the generalization to
model practical systems having safety constraints, design bounds, operational
boundaries, etc. Subsequently, we demonstrated that linear temporal inequalities
with constant coefficients can be formulated using polynomial matrix shift
operators. We analyzed the existence of solutions for such behavioral
inequalities utilizing a generalization of the Farkas' Lemma. Additionally, we
looked at systems where both behavioral equalities and inequalities appear in
conjunction, and derived the conditions for feasibility of such constraints.
Provided solutions to the behavioral inequalities exist, we discussed how to
characterize these solutions using slack variables. Finally, we looked at two
practical examples to test the efficacy of the proposed theory.

The proposed method of behavioral inequalities may play a pivotal role in
developing data-driven constraint aware controllers for safety-critical systems.
In subsequent works, we wish to explore optimization problems with behavioral
inequality constraints. Further research on extending the idea of behavioral
inequalities to continuous-time systems may also be explored.



\bibliographystyle{ieeetr}
\bibliography{sample}

\end{document}